# Micropatterned Charge Heterogeneities via Vapor Deposition of Aminosilanes

Christian Pick[1], Christopher Argento[1], German Drazer*,[2], and Joelle Frechette*,[1]

[1] Department of Chemical and Biomolecular Engineering, Johns Hopkins University, Baltimore MD, 21218, USA, [2] Mechanical and Aerospace Engineering Department, Rutgers University, Piscataway, NJ 08854, USA

*Corresponding authors: email : german.drazer@rutgers.edu, and jfrechette@jhu.edu

**Abstract**

Aminosilanes are routinely employed for charge reversal or to create coupling layers on oxide surfaces. We present a chemical vapor deposition method to pattern mica surfaces with regions of high-quality aminosilane (3-aminopropyltriethoxysilane, APTES) monolayers. The approach relies on the vapor deposition of an aminosilane through a patterned array of through-holes in a PDMS (poly(dimethylsiloxane)) membrane that acts as a mask. In aqueous solutions the surfaces have regular patterns of charge heterogeneities with minimal topographical variations over large areas. This versatile dry lift-off deposition method alleviates issues with multilayer formation and can be used to create charge patterns on curved surfaces. We identify the necessary steps to achieve high quality monolayers and charge reversal of the underlying mica surface: 1) hexane extraction to remove unreacted PDMS oligomers from the membrane that would otherwise deposit on and contaminate the substrate, 2) oxygen plasma treatment of the top of the membrane surfaces to generate a barrier layer that blocks APTES transport through the PDMS, and 3) decrease of the vapor pressure of APTES during deposition to minimize APTES condensation at the mica–membrane-vapor contact lines and to prevent multilayer formation. Under these conditions, AFM imaging shows that the monolayers have a height of (0.9 ± 0.2) nm with an increase in height up to 3nm at the mica–membrane-vapor contact lines. Fluorescence imaging demonstrates pattern fidelity on both flat and curved surfaces, for feature sizes that vary between 6.5-40 μm. We verify charge reversal by measuring the double layer forces between a homogeneous (unpatterned) APTES monolayers and a mica surface in aqueous solution and we characterize the surface potential of APTES monolayers by measuring the double layer forces between identical APTES surfaces. We obtain a surface potential of +110mV +/- 6 mV at pH 4.0.

## 1. Introduction

Spatial control of chemical functionality is critical in the development of platforms for bio-sensing technologies where the localization and immobilization of molecules or particles to surfaces is necessary.[1,2] Of particular interest is the deposition of 3-aminopropyltriethoxysilane (APTES) on oxide surfaces such as $SiO_2$[3,4] or sapphire[4], as well as on mica surfaces[5,6]. APTES contains two different reactive groups: on one end are three ethoxysilanes that can undergo a condensation reaction, covalently attach to surfaces, and crosslink. The other end is a primary amine group that is protonated in aqueous solutions ($pK_A = 9.6$)[7]. Therefore, an APTES-covered surface will be positively charged, allowing for the reversal of the negative charge present on most surfaces in aqueous solutions. Additionally, the primary amine group can undergo further reactions with functional groups such as carboxylic acid, aldehydes, and epoxy groups.[8] This allows APTES films to be used for the covalent attachment of biomolecules onto surfaces[9], making APTES monolayers the foundation layer on many devices.[10,11,12]

Creating high quality APTES monolayers can be a challenge, even without the added difficulties associated with creating microscale patterns. In particular, precursors containing multiple reactive groups, e.g. trichloro- or trialkoxy- functionalities, have condensation reactions that are not self-limiting and therefore films can build up well beyond a monolayer[13]. Control over the deposition conditions is particularly important for organosilanes with a primary amine functional group (such as APTES) because the amine group catalyzes the hydrolysis of alkoxysilane endgroups.[13] As a result, many solution-based deposition procedures of aminosilanes can lead to copolymerization of the precursor molecules in the solution prior to deposition, resulting in the formation of aggregates[4,14] or multi-layers on the surfaces[13]. Conversely, unwanted sub-monolayer coverage of primary amine functional groups on a surface can prevent charge reversal and limit the number of binding sites for covalent attachment of target molecules. While multilayers (or sub-monolayers) may be acceptable in certain applications, they are often undesirable. For example, when used as a coupling layer in biosensing devices, non-uniformities in this coupling layer can adversely affect sensor performance.[9] Similarly, interactions measured with the surface forces apparatus (SFA) are obtained with sub-nanometer resolution in the separation between relatively large surfaces (~1cm$^2$) with mica as the substrate of choice. Therefore functionalization with high quality monolayers on mica[15] is particularly important and a surface patterned with well-defined charge heterogeneities with minimal topographical changes would open the door to study double layer forces between patchy surfaces. The SFA brings about additional requirements for patterning that include the fact that the SFA relies on curved surfaces (radius of curvature of ~2cm).

Both solution and vapor-phase deposition have been employed to create organosilane monolayers.[16] In solution-based deposition, the organosilane precursor is dissolved in a solvent and the surface is subsequently immersed in this solution for a set period of time. One of the primary issues negatively affecting this approach is the undesired deposition of aggregates and multilayers due to the hydrolysis, and subsequent cross-linking of the precursor molecules in solution prior the deposition on the surface.[17] To minimize the formation of these aggregates it is necessary to optimize deposition time, temperature, and organosilane concentration.[13] To limit the amount of dissolved water, anhydrous organic solvents such as toluene[18] or hexane[19] are commonly used. This helps to minimize hydrolysis during silanization[16]. For aminosilanes in particular, much of the previous work on solution-phase deposition has focused on generating high-quality monolayers without the added challenge of creating microscale patterns. Any patterning method should reproducibly yield regions with high-quality aminosilane monolayers while leaving the surrounding area free of any contamination resulting from the patterning process. Microcontact printing[20] is a common solution-based patterning method that has been used for organosilanes[4, 21] and relies on the use of an elastomeric stamp to control the spatial transfer of an "ink" containing the desired species to a surface. While microcontact printing represents a fairly straightforward method of patterning organosilanes onto surfaces, it inherits all the challenges associated with solution-based deposition along with additional ones. For example, variations in contact time and pressure applied to the stamp can result in variability in the patterned layers.[22]

Chemical vapor deposition (CVD) is a practical alternative to create organosilane monolayers. In closed-system vapor deposition, the target surface is placed in an evacuated chamber together with a small dish containing a liquid drop of the organosilane. The organosilane first vaporizes and then condenses on all the surfaces, including the target surface. Some of the advantages of vapor deposition over liquid-based deposition protocols include the reduction in the amount of aggregates on the surface, the elimination of solvents, and a better control over excess humidity during the deposition process.[3, 14, 16, 23] To create a pattern, it is necessary to selectively expose parts of the surface to the vapor phase by using a blocking layer or mask. Alternatively, the deposited organosilane can be selectively desorbed from the surface following deposition. A typical mask consists of a film with open features that are patterned via either e-beam lithography[24] or photolithography[25]. Once the silane deposition is complete, the blocking layer is removed via a lift-off step. It can be challenging, however, to fabricate such a blocking layer on curved surfaces.

Here we show how chemical vapor deposition of APTES monolayers through a PDMS mask can be used to create positively-charge patterns on mica with minimal topographical variations. The method, based on the work of Jackman et al.[26], is relatively simple, relies on the dry lift-off of the PDMS membrane after deposition, leaves the unpatterned surface free of residues, and works on curved surfaces. Our results identify key steps that are essential to yield patterns with good quality monolayers. These steps include hexane extraction and plasma treatment of the membranes, as well as the necessary APTES concentration to minimize topographical variations on the patterned surfaces while maintaining local charge reversal.

## 2. Materials and Methods.

### 2.1 Materials.

Elastomer (Dow Corning Sylgard® 184) is purchased from Robert McKeown Inc. (Branchburg, NJ). SU-8 2025 photoresist and developer are purchased from Microchem Corp. (Newton, MA). 3-aminopropyltriethoxysilane (APTES) 98% and tridecafluoro-1,1,2,2-tetrahydrooctyl trichlorosilane are purchased from Sigma-Aldrich (St. Louis, MO). Mica (Ruby , ASTM V-1/2) is purchased from S&J Trading (Glenn Oaks, NY), and hydrochloric acid (Fisher Chemical, OPTIMA grade) is diluted with deionized water to a concentration of $10^{-4}$ M. Fluorescent carboxylic acid-functionalized particles (diameter = 93nm) are purchased from Bang's Laboratories (Fishers, IN). Unless mentioned otherwise, all chemicals are used as received.

### 2.2 Fabrication

**PDMS membranes.** Molds for the PDMS membranes are fabricated using conventional photolithography. Micropillar arrays, which serve as template for the membrane holes, are fabricated on a silicon wafer using SU-8 2025. Following fabrication, the mold is silanized with tridecafluoro-1,1,2,2-tetrahydrooctyl trichlorosilane for 1 hr at room temperature in a vacuum desiccator. The elastomer base and curing agent are mixed in a 10:1 ratio and degassed under vacuum for 20 minutes. Following degassing, the elastomer mix is spin-coated onto the mold so that the layer deposited is thinner than the height of the micropillars, ensuring the membrane contains through-holes. The final membrane thickness used in this work is 20 μm. Following spin-coating, the elastomer is cured at 70°C for 48 hours to ensure complete cross-linking.[27] Once cured and peeled off the mold, the membranes are imaged under an optical microscope to verify that the pillars are not removed from the mold upon lift-off and that the holes are clean and go through the membrane (see optical micrographs for the membranes in supplemental information Figure S1). Any remaining unreacted PDMS oligomers in the membrane are removed via an overnight extraction in hexanes.[28] Following extraction, the membranes are dried in a vacuum oven overnight at 70°C and cleaned in an ultrasonic bath in 200 proof ethanol 3 times for 5 minutes each. The membranes are then dried again in a vacuum oven at 70°C overnight. Before APTES vapor deposition, the membranes are exposed to an oxygen plasma treatment (50W, 0.3 Torr, and 1 min) on their top surface to produce an oxide layer that acts as a barrier to the transport of small molecules through the membrane, using a home-built plasma reactor. This barrier layer is necessary to reduce the permeability of the membranes to APTES vapor.

**Patterned APTES.** Freshly cleaved mica surfaces are used as the substrates for APTES deposition. Prepared membranes are carefully applied to the mica with tweezers to ensure conformal contact. The mica surfaces covered with the PDMS membranes are placed in a plastic desiccator (Scienceware® vacuum desiccator) that is transferred to a glovebag (Aldrich® Atmosbag). The desiccator is evacuated for 30min with a mechanical vacuum pump, then sealed while the glovebag is purged. The glove bag is purged with high-purity nitrogen 3-5 times to remove traces of moisture. Following purging, the desiccator is opened in the dry nitrogen atmosphere inside the glovebag and a small dish of APTES with a known volume is placed inside. The APTES concentrations reported throughout this work are defined as the APTES drop volume used for the deposition (in microliters) per the internal desiccator volume (in liters). The desiccator is evacuated for 1 min and then sealed to allow silane deposition to occur over a period of 4-12 hours at room temperature (22 °C). We found that 4 hours was the minimum time required for the formation of complete patterns. Following this deposition period, the desiccator is purged with nitrogen and the samples are removed. The PDMS membranes are then lifted off the mica surfaces with tweezers and the surfaces are rinsed with 200 proof ethanol. After the ethanol rinse, the surfaces are dried with filtered nitrogen and ready for subsequent characterization.

**2.3 Characterization.**

**Surface Forces Apparatus (SFA) experiments.** The MK II SFA[29] equipped with microstepping motors is employed to measure the interaction forces between APTES-APTES, APTES-Mica, and Mica-Mica surfaces in aqueous electrolye solutions. In the SFA, the surface separation is estimated from the position of the fringes of equal chromatic order (FECO)[30] resulting from multiple beam interferometry (MBI).[31] The wavelengths at the vertex of the parabolic fringes are used to estimate the surface separation at the point of closet approach for a sphere-plane configuration. To determine surface separation we use the multilayer matrix method[32] combined with the fast spectral correlation algorithm.[33, 34] The interaction between the two crossed-cylinders is calculated from the deflection of a soft cantilever spring ($k = 118.3$ N/m). The radius of curvature, $R = \sqrt{R_1 R_2}$ is determined from the geometric mean of two spatially resolved FECO profiles coming from perpendicular cross-sections.

<u>Cleaning.</u> All stainless steel parts that come into contact with electrolyte (spring, upper, and lower disk holder) are cleaned in an RBS 35 (Pierce, Rockford, IL) detergent solution, passivated in 50% nitric acid, rinsed thoroughly with ethanol, and dried immediately before use. All of the Teflon parts (bath, tubing assembly) are cleaned in a detergent solution, rinsed thoroughly with water, and dried with nitrogen immediately before use. All glassware is cleaned with detergent, and rinsed with water.

<u>Surface preparation.</u> For the surfaces used in in the SFA, 3-5 μm thick mica pieces are cleaved in a laminar hood and placed on a larger backing sheet. The cleaved mica pieces are coated with 50 nm of silver (99.999% purity, Alfa Aesar) via thermal evaporation (Kurt J. Lesker Nano 38) at a rate of 2-3 Å/s. The mica pieces are then glued (on the silvered side) onto a silica support disk for the SFA. For the APTES deposition, the entire disk/silvered mica combination is placed inside the vacuum desiccator and transferred to a glovebag. The APTES deposition procedure follows the same protocol for the patterned surfaces.

Procedure. A Teflon bath is employed inside the SFA chamber and 25 mL of the electrolyte solution is injected while the surfaces are separated using a syringe equipped with all Teflon tubing and valves. The solution is left in the apparatus for 1-2 h for equilibration prior to force measurements. Each force profile (approach and retraction) is repeated at least 5 times. All experiments were performed at 23 °C.

Double layer interactions. Measurement of double layer forces and their comparison with DLVO (Derjaguin-Landau-Verwey-Overbeck) theory is employed to determine the surface potential and surface charge density of the APTES-covered surfaces. DLVO theory[35, 36, 37] describes the interaction between two flat surfaces in an electrolyte solution as the superposition of the van der Waals and electrostatic interaction energies. We calculate the electrostatic interaction energy from the excess pressure in the gap, calculated by solving numerically the full Poisson-Boltzmann equation for both constant potential and constant charge boundary conditions using MATLAB's boundary value problem solver (bvp5c), and the electrostatic interaction energy is obtained from a numerical integration of the pressure (see details for the numerical method in the supplemental information). Hamaker theory is used for the non-retarded van der Waals interactions with a Hamaker constant of $2.2 \times 10^{-20} J^{38}$ for the interactions between mica surfaces in aqueous solutions. Finally, we employ the Derjaguin approximation to convert the interaction energy between flat surfaces to the forces normalized by the radius of curvature between crossed-cylinders. In comparing to DLVO theory, the measured forces were fitted for both a Debye length and surface potentials. The fitted Debye length was obtained from a least-squares fit of the force data to an exponential function at separations greater than 1 expected Debye length, $\kappa^{-1}$. The Debye length is calculated for a 1-1 electrolyte using: $\kappa^{-1} = \sqrt{\varepsilon_0 \varepsilon_r kT / 2e^2 n_b}$, where $n_b$ is the bulk ion concentration, k is Boltzmann constant, T is temperature, $\varepsilon_0$ is the permittivity of free space, $\varepsilon_r$ is the relative permittivity of the solution. The surface potential and charge density of the surfaces were obtained from a least squares fit of the data to predictions for both the constant charge and constant potential boundary conditions.

**Fluorescence Imaging.** Surfaces are tagged by soaking them for 30-45 minutes in a $10^{-5}$ volume fraction solution of carboxyl-functionalized fluorescent particles dispersed in deionized water (18.2 MΩ·cm). Following soaking, the surfaces are rinsed with deionized water and dried with nitrogen. Fluorescence images of the tagged surfaces are taken with an Olympus BH-2 microscope equipped with a Tucsen 3.3MP CCD camera. TSview version 6 is used for image capture. Pattern dimensions and area coverage are measured using ImageJ 1.46r. Coverage is determined by converting the fluorescent images to binary format in ImageJ and measuring the area coverage using the built-in particle analyzer.

**Atomic force microscopy (AFM) imaging.** Topographical and phase images of the patterned APTES monolayers are taken with a Bruker Dimension 3100 AFM in tapping mode with a scan rate of 1.5 Hz and a scan area of 50x50 µm. The height of the APTES layers is measured in Bruker Nanoscope Analysis version 1.40 after performing a third order flattening of the raw height images. Figures 4 and S5 are made

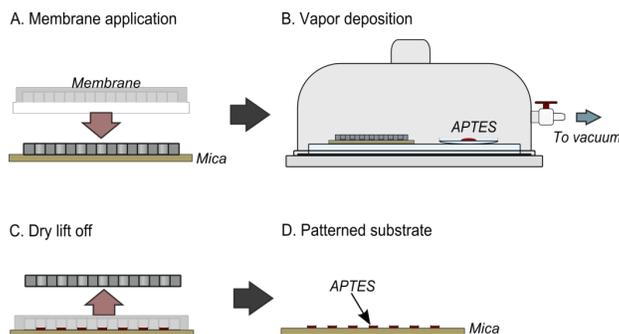

**Figure 1.** Diagram of patterning CVD steps. A) A plasma treated and hexane extracted PDMS membrane is placed on a mica surface (plasma treated side facing up). B) the mica surface is placed in a partially evacuated desiccator in the presence of an APTES drop and left to react for 4-12 hours. C) After the deposition, the membrane is lifted from the surface to yield (D) patterned areas of APTES monolayers.

in Gwyddion 2.35.

## 3. Results and Discussion.

### *3.1 CVD deposition of patterned APTES monolayers.*

The patterning procedure extends the work of Jackman et al.[26] to aminosilane monolayers. The chemical vapor deposition of APTES relies on the condensation of the molecules from the vapor phase on the accessible areas of a mica surface. Arrays of APTES monolayers are formed by blocking part of the mica substrate with a PDMS membrane that has been patterned with a hexagonal array of through-holes (see Fig. 1 and supplemental information Fig. S1 for optical micrographs of the membranes). In principle, APTES deposition only occurs through the membrane holes, and the PDMS membrane acts as a mask (Fig. 1B-D). The APTES vapor comes from a droplet of APTES of known volume allowed to evaporate in a partially evacuated desiccator (Fig. 1B). The volume of the drop controls the concentration (partial pressure) of APTES in the vapor phase, and needs to be optimized to yield high quality monolayers. The APTES concentration is defined as the APTES drop volume (in microliters) per internal desiccator volume (in liters). Advantages of this method for patterning APTES are the dry lift-off, i.e. it is resist-free (does not require the chemical removal of a sacrificial layer), and the mechanical flexibility of the membrane allowing for the patterning on curved surfaces. In developing the process we faced two important challenges unique to working with a PDMS membrane as a mask: 1) the inherent permeability of PDMS to small molecules[22], and 2) the transfer of PDMS oligomers from the membrane to the mica surface[27, 28, 39]. Additional challenges associated with the deposition process include achieving a good pattern fidelity over large areas and making high quality monolayers with mimimal topographical heterogenities.

Transport of APTES through the PDMS membranes can lead to its deposition outside of the desired patterns (in the areas blocked by the membrane). Using thicker membranes and shorter deposition times can help reduce some of the APTES transport through the membrane material. However, we found that plasma treatment of the membranes, prior to their contact with the mica substrates, blocks the diffusion of APTES through the membranes and prevents deposition outside of the open areas. Plasma treatment of PDMS is known to form a silica-like oxide layer on the PDMS surface[40, 41], which has been reported to hinder the diffusive transport of small molecules through bulk PDMS[42]. To characterize the effectiveness of the plasma treatment, we tagged patterned mica surfaces with negatively charged fluorescent particles to determine the extent of APTES deposition outside of the patterned areas. In the absence of plasma treatment we observe particle deposition everywhere on the mica surface (see fluorescence images in Fig. S2 of the supplemental information). In constrast, we do not observe particle deposition outside of the patterned areas when the top of the PDMS surface has been exposed to oxygen plasma. The right conditions for the plasma treatment are critical to its success in blocking APTES diffusion. If the plasma treatment is too long, cracks appear on the PDMS surface (see optical micrographs in Fig. S3 of the supplemental information). The presence of cracks increases the permeability of the PDMS to the APTES molecules and leads to deposition outside of the desired areas. On the other hand, if the plasma treatment is too short the oxide layer formed is not sufficient to prevent APTES diffusion. We found that a 300 mTorr and 50W oxygen plasma treatment for 1 minute worked best. Although plasma treatment performed on the side of the PDMS membrane that is in contact with the mica surface was also found to prevent diffusion of APTES through the membrane, it significantly increases the adhesion between the mica and the PDMS membrane. This increase in adhesion makes lift-off difficult and can even leave pieces of PDMS on the mica. Therefore, we opted to perform the plasma treatment on the top-side of the PDMS membranes (the side exposed to the APTES vapor).

Cured PDMS is known to contain traces of unreacted oligomers[28] that can be transferred to the underlying mica substrate, leaving unwanted residues on the surface after the membrane is lifted-off. We investigated if extended curing of the PDMS membrane followed by hexane extraction[28] (see methods section) could significantly reduce transfer of oligomers to the mica surface. We performed complete APTES deposition procedures on mica surfaces covered with PDMS sheets of the same thickness as the patterned membranes. We considered both *extracted* and *unextracted* PDMS sheets. After the APTES deposition and membrane lift-off we measured the double layer forces in $10^{-4}$M HCl (pH 4.0) using the SFA, see Fig. 2. Mica surfaces that have been in contact with unextracted PDMS sheets show strong short-range repulsive forces that cannot be described by DLVO theory alone. We attribute these forces to the transfer of reacted and unreacted PDMS to the mica (Figure 2A). In contrast, no short range steric forces are observed for the mica surfaces that have been in contact with the extracted PDMS sheets (Figure 2B). In this case, the surfaces jump into van der Waals contact. In addition, the surface forces between these mica surfaces are well-described by DLVO theory with surface potentials in agreement with those obtained for fresh mica surfaces (see for comparison values reported in Table 3 and the force curves in Fig. S4 of the supplemental information). Therefore based on these results we find that hexane extraction reduces unwanted transfer of the membrane material to the mica surface. Note here that to act as a true control experiment the PDMS sheets remained in contact with the mica surface for as long as the APTES deposition step, and 5 μL/L of APTES vapor was also present in the chamber for the whole process. Due to its positive charge, partial APTES deposition through the membrane would have rendered the surface potential of the mica surface less negative, a feature we do not observe here. Additionally, the sign of the surface potentials were verified by attempting to tag the mica surfaces after the SFA experiments with negatively charged fluorescent particles. No particle deposition was observed on the surfaces, indicating that a net negative surface potential is maintained.

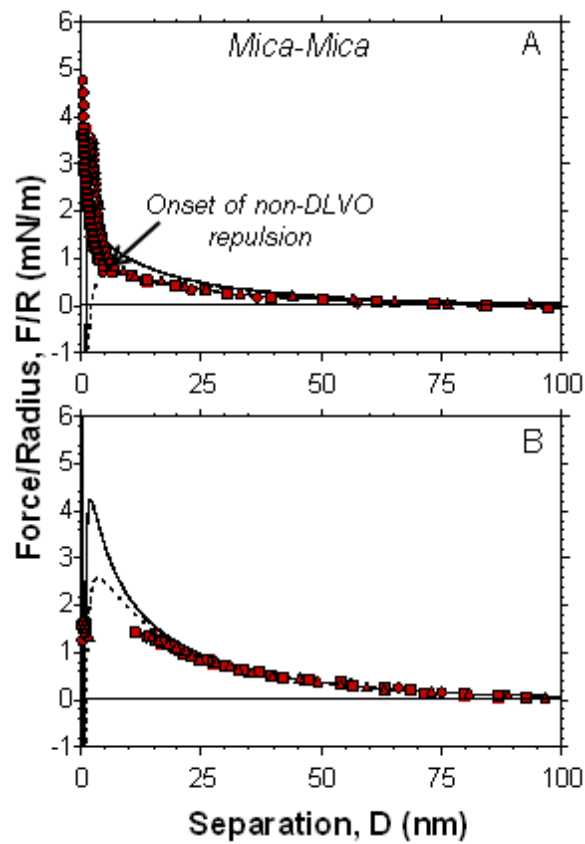

**Figure 2.** Measured force (normalized by the radius of curvature) in $10^{-4}$ M HCl solution between two mica surfaces as a function of surface separation. Prior to force measurements the mica surfaces were in contact with PDMS sheets with APTES vapor present. The PDMS sheet was A) unextracted with a barrier layer generated on the bottom, and B) extracted membrane with a plasma-generated barrier layer on the top. Solid lines represent DLVO fits with constant charge boundary conditions and dashed lines represent constant potential boundary condition.

We investigate the effect of both APTES concentration during deposition and feature sizes on pattern fidelity by tagging the surfaces with negatively charged fluorescent particles (see Figure 3). These particles deposit on the positively charged (APTES) regions of the surface but not on the negative ones (bare mica) allowing us to quantify APTES pattern fidelity using image analysis. Table 1 compares the size of the APTES features, as determined from fluorescent images, to the ones expected based on the array dimensions on the photomask used to fabricate the membrane template. We obtain good pattern fidelity for all the APTES concentrations investigated here and find that the APTES concentration in the desiccator has no measurable effect on the overall quality of the patterns. For patterns created on flat surfaces with feature sizes that are greater than 30 microns we observe that the diameter of individual patterned circles is uniform over large areas of the surface (>1cm$^2$) and in excellent agreement with the expected values. However, for curved surfaces and for smaller features (6.5 µm x 6.5 µm), we find that, although we retain the hexagonal array and the pattern features are very uniform over large area, the diameter of the patterned circles is consistently significantly smaller than expected. We suspect that the discrepancy is due to the large aspect ratio of the holes in the membranes when patterning smaller features which could hinder transport of APTES. For example, for the features shown in Figure 3C the hole diameter is 6.5 µm and the membrane thickness is 20 µm. In addition, deformation of the membranes when they are in contact with the substrates could alter the size and shape of the patterned features, especially for curved surfaces since the membranes must deform to conform to the curvature.

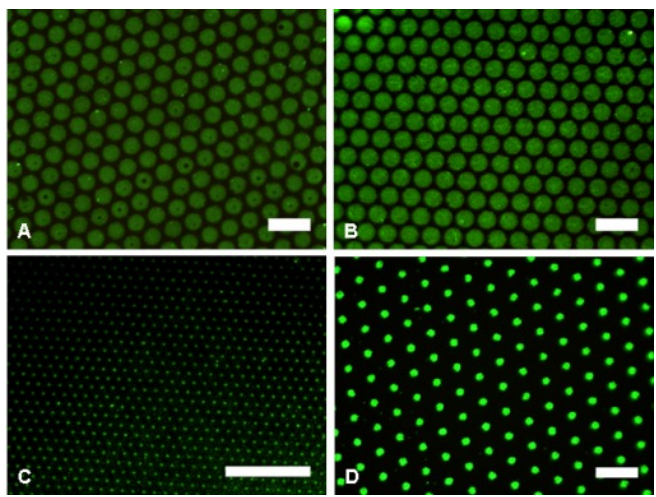

**Figure 3.** Optical micrographs of APTES patterned mica tagged with fluorescent particles The following APTES deposition concentrations were used: A) 0.25 µL/L, B) 1.25 µL/L, C) 5 µL/L, D) 1.25 µL/L (curved surface, ~2cm radius). Scale bar = 100 µm. Dimensions of the patterned features are given in Table 1.

**Table 1.** Measured pattern dimensions from image analysis and comparison to predicted values.

| APTES concentration (μL/L) | Array dimensions diameter x spacing (μm) | Measured APTES patch diameter (μm) | Expected area coverage (%) | Measured area coverage (%) |
|---|---|---|---|---|
| 0.25 (Fig. 3A) | 40 x 12 | 37.2 ± 0.3 | 53.7 | 51.5 ± 0.7 |
| 1.25 (Fig. 3B) | 40 x 12 | 38.0 ± 1.1 | 53.7 | 53.4 ± 2.8 |
| 5.00 (Fig. 3C) | 6.5 x 6.5 | 3.6 ± 0.1 | 21.7 | 9.5 ± 0.2 |
| 5.00 (not shown) | 30 x 9 | 29.0 ± 1.5 | 53.7 | 47.7 ± 4.3 |
| 1.25[a] (Fig. 3D) | 40 x 40 | 24.1 ± 1.1 | 21.7 | 9.4 ± 0.9 |

[a] curved surface

While tagging the patterned areas on the surfaces with fluorescent particles showcases pattern fidelity over large areas, it does not allow us to determine the quality of the APTES monolayers within the deposited areas. We characterize the quality of the monolayers within an individual patterned circle using AFM (Fig. 4 and Table 2, as well as higher resolution images in Fig. S5 of the supplemental information). AFM imaging can determine the height of the monolayers and identify the presence of aggregates or multilayers on the surfaces. For the three different APTES deposition concentrations investigated, the average height of individual feature are uniform and all at least 0.8nm, in agreement with reported values for a full monolayer[10]. Moreover we do not see evidence of large APTES aggregates inside the patterned areas. However, the APTES height observed for the 5 μL/L concentration is nearly twice the reported value for a monolayer (Table 2), indicating nearly a bilayer coverage. Therefore this concentration should be avoided if a monolayer deposition is required.

The AFM images also indicate the presence of thicker rings around the patterned APTES features, see Figure 4C and a higher resolution image for the 0.25 μL/L concentration in Figure S5) This is particularly noticeable in the case of the 5 μL/L (Fig. 4C). We suspect that the rings are due to the condensation of the APTES at the triple contact line, which is where PDMS, APTES condensate (and residual water condensate), and mica meet. Similar raised edges have been observed during vapor deposition within PDMS microchannels by George et al.[43] Capillary condensation at the triple contact line is a barrierless nucleation process for unsaturated vapors that is described by the Kelvin equation (Eqn 1)[44]

$$\ln\left(\frac{P}{P_{sat}}\right) = \frac{2\gamma V_M}{rRT} ,  \quad (0)$$

where 1/r is the meniscus curvature, which is negative for a concave meniscus. Here we approximate the the height of the condensate as –r. $P/P_{sat}$ is the partial pressure with $P_{sat}$ = 10 Pa at 22°C[45], $\gamma$ is the APTES surface tension (assumed equal to 21 mN/m which is a reported value for triethoxysilane at 20°C)[46], $V_M$ is the molar volume of APTES, R is the ideal gas constant, and T is the temperature. Unsaturated conditions ($P<P_{sat}$) should decrease the ring height but cannot completely eliminate the rings. We estimate that a drop volume of about 1.0 μL/L or greater results in saturated conditions if we assume that the entire drop evaporates until saturation is reached. Therefore, the APTES drop volume of 0.25 μL/L corresponds to unsaturated conditions, while there is sufficient APTES in the 5 μL/L concentration to reach saturation (see Table 2). The 1.25 μL/L drop is estimated to generate a pressure around the saturated limit. To minimize the condensation ring height, it is important to optimize the amount of APTES used in the chamber: we need a monolayer coverage but rings that are as small as possible. Based on the AFM and fluorescent imaging, a concentration of 0.25 μL/L minimizes the ring height to that of a bilayer while maintaining a monolayer coverage on the rest of the patches and excellent patterned fidelity.

Table 2. AFM height data for patterned surfaces.

| APTES concentration (μL/L) | $P/P_{sat}$ | APTES Height (nm) | Condensation Ring height (nm) | Predicted Kelvin radius (nm) |
|---|---|---|---|---|
| 5 | 5.29 (*saturated*) | 1.4 ± 0.4 | 12.8 ± 2.0 | N/A |
| 1.25 | 1.32 (*saturated*) | 0.9 ± 0.2 | 3.1 ± 0.8 | N/A |
| 0.25 | 0.26 | 0.8 ± 0.1 | 1.6 ± 0.8 | 3.1 |

### 3.2 Charge density of APTES monolayers.

We measured the surface forces between two APTES monolayers in $10^{-4}$ HCl aqueous solution to determine how the APTES concentration during deposition influences the surface potential and robustness of the monolayers. Surface forces measured in an aqueous electrolyte solution between ideal APTES monolayers would be well-described by DLVO theory, have a positive surface potential, and be reproducible over multiple approach and retraction curves. Shown in Figure 5A-C are the surface forces for APTES films deposited using the three different APTES concentrations. Note that here the APTES deposition on mica is performed without a PDMS membrane (no patterns). The lines in Fig. 5A-C represent predictions from DLVO theory[36, 37] that were fitted to obtain the surface potential of APTES ($\psi_{APTES}$) and the Debye length. In all cases the APTES layers appear stable and robust and we see that the measured forces are reproducible over multiple approach and retraction curves, even after repeated contact and pull-out cycles. For separations greater than a Debye length, the forces between APTES monolayers are well-described by DLVO theory for the three APTES concentrations investigated here and are better described by the constant charge boundary condition (solid line). The decay length of the forces is also in good agreement with predictions for a $10^{-4}$M (1-1)electrolyte concentration (31nm). Forces between APTES monolayers formed at the highest APTES concentration (5 µL/L) during deposition display repulsion close to contact that is not accounted for by DLVO theory (Fig. 5A). This additional steric repulsion could be explained by the existence of multilayers on the surface, which is consistent with the AFM measurements shown in Fig. 4C. This additional repulsion prevents the surfaces prepared at the highest concentration (5 µL/L) from reaching adhesive contact, which is in contrast with monolayers prepared at the two lowest concentrations (see the pull-out forces in Table 3).

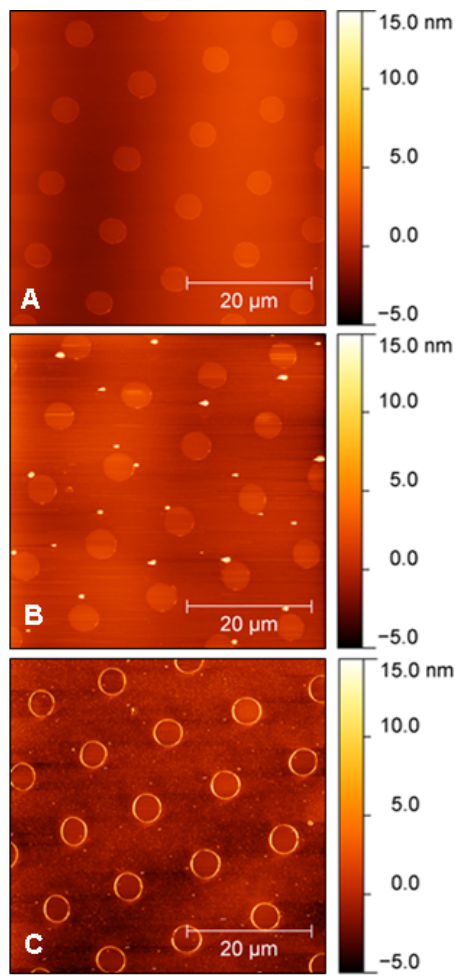

**Figure 4.** AFM height images of patterns deposited at different APTES concentrations A) 0.25 µL/L B) 1.25 µL/L C) 5 µL/L

Comparison between the measured forces and DLVO theory for two identical surfaces gives the magnitude but not the sign of the surface potential. To determine the sign of the surface potential of the APTES surfaces, we measured the double layer forces between a bare mica surface (known negative surface potential) and APTES-covered mica surfaces prepared under identical conditions as in Fig. 5A-C (see Fig. 5D-F). The surface potential of mica is well-characterized in the literature[38] and has been measured separately (see Fig. 2B, and Fig. S4 in supplemental information). Here we use $\psi_{mica} = -120 mV$. Also shown in Fig. 5 D-F are DLVO predictions for the asymmetrical interactions calculated based on the value of $\psi_{APTES}$ obtained from the corresponding APTES-APTES covered surfaces assuming that the value of $\psi_{APTES}$ obtained in Fig. 5 A-C is positive. By comparing the symmetric (APTES-APTES) and asymmetric (APTES-mica) interactions, we find that we achieve charge reversal only at the two highest APTES concentrations (See Table 3 for fitted surface potentials and corresponding surface charge densities). The repulsive interaction for the APTES-mica forces shown in Fig. 5F indicates that the 0.25 µL/L APTES surface potential is negative. The absence of charge reversal is indicative of an incomplete APTES monolayer on the mica surface, even if the height of the monolayer as measured in the AFM indicates a full monolayer. This discrepancy between the SFA force measurements and the AFM height data is surprising, but could be related to the capillary condensation observed to produce rings shown in Fig. 4. Condensation at the triple contact line, as observed with the AFM, provides a nucleation site for the APTES vapor which facilitates monolayer formation. In contrast, no membranes are used to create unpatterned monolayers for the force measurements, therefore these sites for nucleation at the triple contact line are absent. This discrepancy is likely less important at higher APTES concentrations where $P_{APTES}=P_{sat}$ and condensation can occur everywhere.

The mica/APTES pull-off forces also increases with APTES concentration during deposition (Table 3), which is consistent with having more APTES on the surface. We also observed that the pull-off forces for the 0.25 µL/L APTES symmetric is quite large, and similar to APTES-mica pull-out forces. This large adhesive forces was reproducible over multiple approach and retraction cycles. We suspect that it might be due to incomplete APTES monolayers present on both surfaces where, for example, APTES domains and bare mica interact in contact leading to large adhesion forces.

**Table 3.** Fitted values for DLVO theory for the forces curves in Fig 2 and Fig. 5. The expected Debye length at $10^{-4}$M and 23°C = 30.6 nm

| APTES concentration (µL/L) | Debye length, $\kappa^{-1}$ (nm) | Surface potential, $\Psi_s$ (mV) | Surface charge density, $\sigma$ (e/nm$^2$) | $-F_{adh}/R$ (APTES-mica) mN/m | $-F_{adh}/R$ (symmetric) mN/m |
|---|---|---|---|---|---|
| 5.00 | 30.9 ± 2.4 | 117 ± 9 | 0.036±0.006 | 122.4 ± 7.0 | 0 |
| 1.25 | 30.0 ± 0.8 | 110 ± 6 | 0.032±0.004 | 77.1 ± 8.0 | 1.5 ± 0.9 |
| 0.25 | 31.5 ± 3.8 | -99 ± 11 | -0.025±0.006 | 12.3 ± 0.5 | 134.4 ± 6.0 |
| Extracted:mica | 30.9 ± 0.8 | -120 ± 5 | -0.038±0.004 | N/A | 28.9 ± 2.3 |
| Unextrated:mica | 31.5 ± 6.5 | -74 ± 3 | -0.015±0.001 | N/A | 0 |

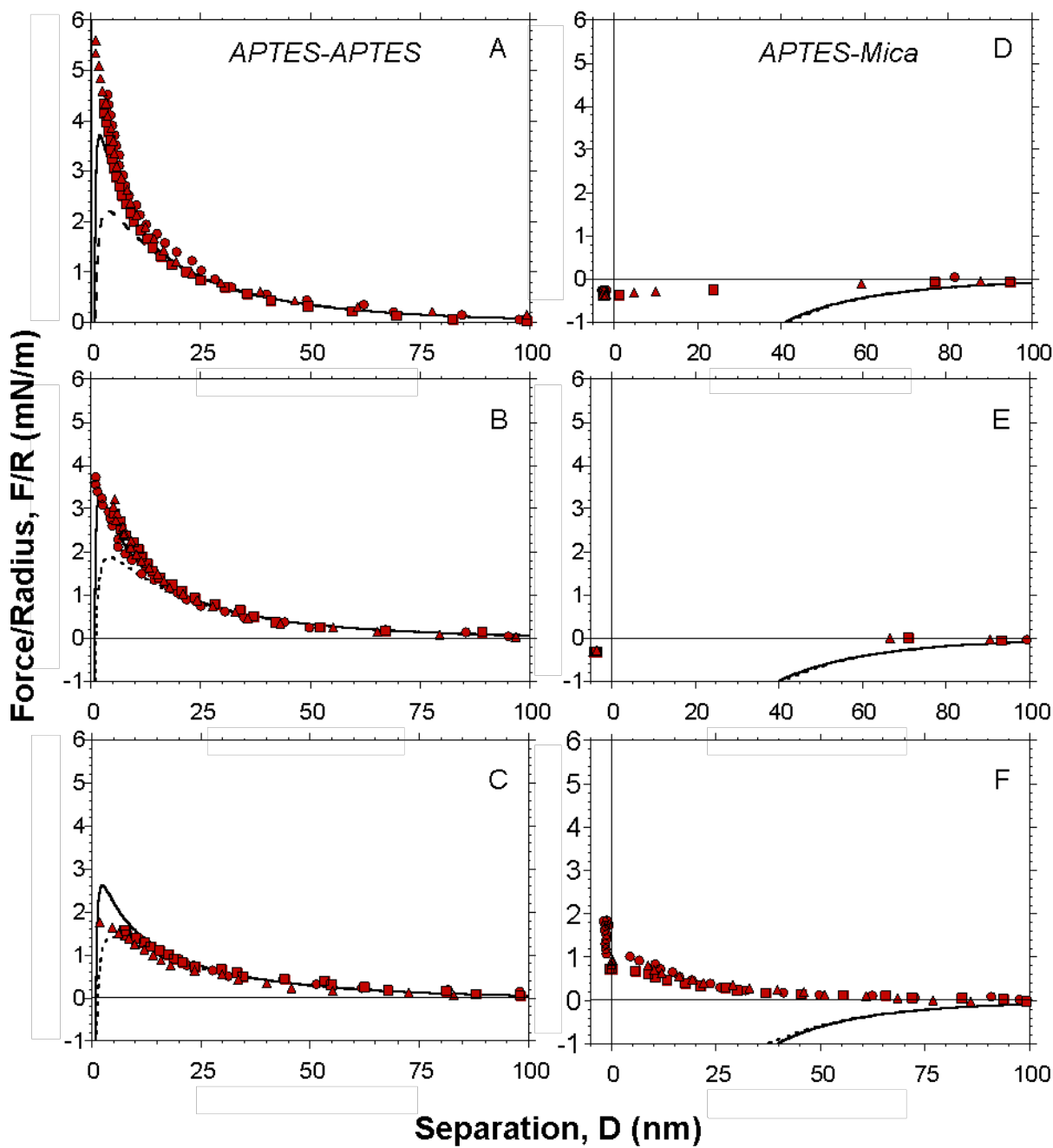

**Figure 5.** Force curves, normalized by the radius of curvature, as a function of separation measured between (A-C) APTES-APTES (symmetric), and (D-F) APTES-mica (asymmetric) surfaces in $10^{-4}$M HCl. Each plot show multiple approach/retraction curves, each indicated by different symbols. The APTES concentration during the deposition are (A,D) 5 µL/L, (B,E) 1.25 µL/L, (C,F) 0.25 µL/L. Solid lines represent DLVO fits with constant charge boundary conditions and dashed lines represent constant potential boundary condition. The sign of $\psi_{APTES}$ in (A-C) is determined from the force measurements in (D-F).

While APTES monolayers are used extensively to reverse the charge of a negatively charged surface, the surface potential of a monolayer of APTES has not been characterized extensively through force measurements. Grabbe measured the surface forces between films of (γ-aminopropyl)-dimethylethoxysilane (APDMS) on silica surfaces using the SFA under similar conditions as in our experiments ($10^{-4}$ M NaCl solution at pH 5.15)[47]. Agreement with DLVO theory was found for a surface potential of +24.6mV. This surface potential is significantly lower than the value we obtain based in our surface force measurements with APTES monolayers. It is likely that the discrepancy arises because APDMS contains a single ethoxy group to bind to a surface while APTES has three. In contrast to APTES, APDMS molecules are unable to crosslink with other molecules on the surface, which can prevent the formation a dense monolayer on the surface.

More extensive information is available from electrokinetic measurements of APTES functionalized surfaces. For example, Lin et al.[48] conducted streaming potential measurements of APTES deposited on glass from an acetone solution in 1mM NaCl at different pH values. For a pH value of ~4.0, they obtain a streaming potential of about +92mV. Similarly, Na et al. reported a zeta potential value of +93.8mV in 1mM NaCl for a vapor deposited APTES layer on glass.[25] A similar aminosilane, 3-aminopropyltrimethoxysilane (APTMS), deposited on silicon wafers was also reported to have a streaming potential of about +92mV at pH 4.0.[49] These values are in agreement with our measured potentials for APTES films ($\psi_{APTES}$ = +110 ± 6 mV for the 1.25 μL/L APTES concentration), especially when considering that streaming potential values are expected to be lower than surface potential values obtained through direct force measurement, as they are measured at the slip plane away from the surface.

## 4. Conclusions

We have shown a method that relies on elastomeric membranes to spatially control the chemical vapor deposition of high quality aminosilane monolayers to create microscale charge heterogeneities on mica substrates. The advantages of the method include a dry lift-off of the elastomeric membranes that leaves the unpatterned areas free of residues, the capability to pattern on curved surfaces, a high pattern fidelity of full monolayers with minimal topographical variation at the nanoscale, and the absence of aggregates on the surface. The surface potential of both the APTES films and mica were obtained from a series of direct force measurements. These direct force measurements indicate the deposition conditions necessary for charge inversion of the underlying mica surfaces. Additionally, pattern fidelity was characterized by tagging the APTES patterns with fluorescent particles and observing the patterns under a fluorescence microscope. Finally, APTES height measurments were taken with an AFM revealed the formation of condensation rings when the APTES vapor pressure is close to saturation. Additionally, we found that a hexane extraction and plasma treatment of the PDMS membranes were necessary to prevent oligomer contamination of mica from the membranes and to block APTES diffusion in the PDMS, respectively. Ultimately, we find that 1.25μL/L is the optimum concentration for the reproducible deposition of APTES monolayers on mica, as it leads to charge reversal (Fig. 5 and Table 3), good pattern fidelity (Fig. 3), a full monolayer (Fig. 4), and low condensation ring height (Table 2).

## 5. Acknowledgements

This work is partially supported by the National Science Foundation (NSF-CBET ENV 1436482 and NSF-CBET IPT 0933605) and by the Donors of the American Chemical Society Petroleum Research Fund under Grant 51803-ND5. We are particularly grateful to Patricia McGuiggan for performing the AFM measurements and reviewing the manuscript, and we would like to thank Markus Valtiner for initial guidance with the CVD protocol.